\documentclass[runningheads]{llncs}
\usepackage{caption}
\usepackage{subcaption}
\usepackage{booktabs}
\usepackage{multirow}
\usepackage{graphicx}
\usepackage{hyperref}
\usepackage[subtle,margins=normal,leading=normal]{savetrees}

\usepackage[T1]{fontenc}
\usepackage{graphicx}

\begin{document}
\title{Model Ensemble for Brain Tumor Segmentation in Magnetic Resonance Imaging}
\titlerunning{Brain Tumor Segmentation in MRI}
%

\author{Daniel Capell\'{a}n-Mart\'{i}n\inst{1,2}*, Zhifan Jiang\inst{1}*, Abhijeet Parida\inst{1}*, Xinyang Liu\inst{1}, Van Lam\inst{1}, Hareem Nisar\inst{1}, Austin Tapp\inst{1}, Sarah Elsharkawi\inst{1,3}, Mar\'{i}a J. Ledesma-Carbayo\inst{2}, Syed Muhammad Anwar\inst{1,4}  and Marius George Linguraru\inst{1,4}}

%
\authorrunning{D. Capell\'{a}n-Mart\'{i}n, Z. Jiang, A. Parida et al.}
%
\institute{Sheikh Zayed Institute for Pediatric Surgical Innovation, Children’s National Hospital, Washington, DC 20010, USA \and Biomedical Image Technologies, ETSI Telecomunicaci\'{o}n, Universidad Polit\'{e}cnica de Madrid, Madrid 28040, Spain and CIBER-BBN, ISCIII, Madrid, Spain \and Princeton University, Princeton, NJ 08544, USA \and School of Medicine and Health Sciences, George Washington University, Washington, DC 20052, USA}

%

\maketitle              
\begin{abstract}
Segmenting brain tumors in multi-parametric magnetic resonance imaging enables performing quantitative analysis in support of clinical trials and personalized patient care. This analysis provides the potential to impact clinical decision-making processes, including diagnosis and prognosis. In 2023, the well-established Brain Tumor Segmentation (BraTS) challenge presented a substantial expansion with eight tasks and 4,500 brain tumor cases. In this paper, we present a deep learning-based ensemble strategy that is evaluated for newly included tumor cases in three tasks: pediatric brain tumors (PED), intracranial meningioma (MEN), and brain metastases (MET). In particular, we ensemble outputs from state-of-the-art nnU-Net and Swin UNETR models on a region-wise basis. Furthermore, we implemented a targeted post-processing strategy based on a cross-validated threshold search to improve the segmentation results for tumor sub-regions. The evaluation of our proposed method on unseen test cases for the three tasks resulted in lesion-wise Dice scores for PED: 0.653, 0.809, 0.826; MEN: 0.876, 0.867, 0.849; and MET: 0.555, 0.6, 0.58; for the enhancing tumor, tumor core, and whole tumor, respectively. Our method was ranked first for PED, third for MEN, and fourth for MET, respectively.


* These authors contributed equally.

\end{abstract}
\section{Introduction}
The brain tumor segmentation (BraTS) challenge held in conjunction with the International Conference on Medical Image Computing and Computer Assisted Intervention (MICCAI) conference, established in 2012, has generated a benchmark dataset for the segmentation of adult brain gliomas \cite{brats2021,bakas1,bakas2,bakas3,brats2015}. The BraTS 2023 challenge has expanded to a cluster of challenges, encompassing a variety of tumor types alongside augmentation tasks \cite{brats-ssa2023,anahita2023,medperf,brats-syn2023,li2023brain,Rohlfing2010}. Herein, we propose a segmentation technique for newly introduced tasks featuring smaller datasets or new types of tumors. Particularly, these tasks include pediatric brain tumors (PED) \cite{PEDarxiv}, intracranial meningioma (MEN) \cite{MENarxiv}, and brain metastasis (MET) \cite{METarxiv}. This paper presents the methodology primarily developed for PED tumor segmentation, which was also adapted for MEN and MET segmentation tasks.

Brain cancer has become the leading cause of cancer death among children in the United States \cite{Curin2016}. Although rare, pediatric high-grade brain tumors can be aggressive. For example, the median overall survival was reported to be less than one year for pediatric diffuse mid-line gliomas (DMGs, which replaced the formerly known diffuse intrinsic pontine gliomas (DIPGs) to emphasize that the disease may affect areas other than the pons) \cite{2docm}. Multi-parametric magnetic resonance imaging (mpMRI) is essential in pediatric brain tumor diagnosis and monitoring of tumor progression. While adult and pediatric brain tumors share certain similarities, their locations in the brain and imaging characteristics may be different. For DMGs, necrosis is rare or unclear and the tumor may or may not present enhancement on post-gadolinium T1-weighted MRI. Hence, imaging tools specially designed to analyze pediatric brain tumors are necessary to improve clinical management. Automatic tumor segmentation is usually the first and most important step for the success of such analysis. Over the years, several methods have been presented for adult brain tumor segmentation using the BraTS dataset, however, efforts are needed to develop methods more suitable for pediatric brain tumors. The recently introduced PED task in BraTS 2023 provided an opportunity to develop such methods. 

Although, a few previous works addressed pediatric brain tumors \cite{anahita2023,nf1_opg_seg,4docm}, their automatic segmentation remains challenging. This is due to a lack of training data and variation in heterogeneous histologic sub-regions including peritumoral edematous/invaded tissue, necrotic core, and enhancing tumor. BraTS-PED 2023 provides these types of data  for the first benchmarking initiative on pediatric brain tumor segmentation. The task provides the largest annotated publicly-available retrospective cohort of high-grade gliomas including astrocytomas and DMG/DIPG in children. 

In this work, we developed an ensemble approach involving two state-of-the-art deep learning models. Our approach to segmenting pediatric tumors is tested on two additional tasks at BraTS 2023: the segmentation of meningiomas (MEN) and brain metastases (MET). Meningioma is the most common primary intracranial tumor in adults and can result in significant morbidity and mortality for affected patients. Brain metastases are the most common form of central nervous system (CNS) malignancy in adults. Accurate detection of small metastatic lesions is essential for patient prognosis, as missing even one lesion can lead to repeated interventions and treatment delays. For each task, training was performed only on the dataset provided by each sub-challenge.

\section{Methods}

\begin{figure}[htbp]
  \includegraphics[width=\textwidth]{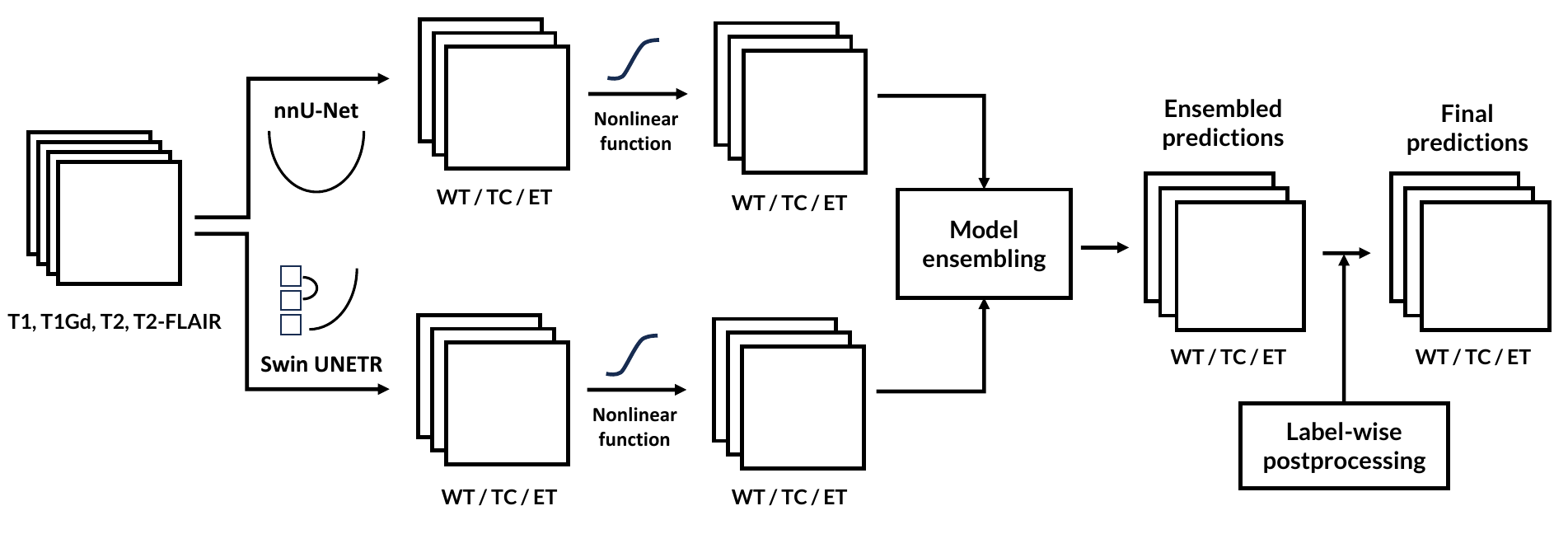}
  \caption{Proposed method: model ensembling and post-processing pipeline. Outputs are obtained from two state-of-the-art deep learning models. These outputs are subjected to nonlinear activation functions and ensembling strategies. Finally, the ensembled predictions are subjected to a specifically tailored post-processing step.}
  \label{fig:pipeline}
\end{figure}

The model ensemble technique is a widely recognized strategy used with machine learning methods and is aimed at increasing the stability and accuracy of model predictions. To harness the benefits inherent to both convolutional neural networks and vision transformers, we adopted an ensemble approach (see Fig. \ref{fig:pipeline}) involving two state-of-the-art models: nnU-Net (“no new U-Net”, winner of BraTS 2020) \cite{nnunet} and Swin UNETR (“Swin U-Net transformers”, top-performing model of BraTS 2021) \cite{swinunetr}. Given that all tasks address the segmentation of multiple tumor sub-regions, our analysis and ensemble approach were conducted in a label-wise manner.

\subsection{Data description}

\begin{figure}[htbp]
    \centering
  \includegraphics[width=0.9\textwidth]{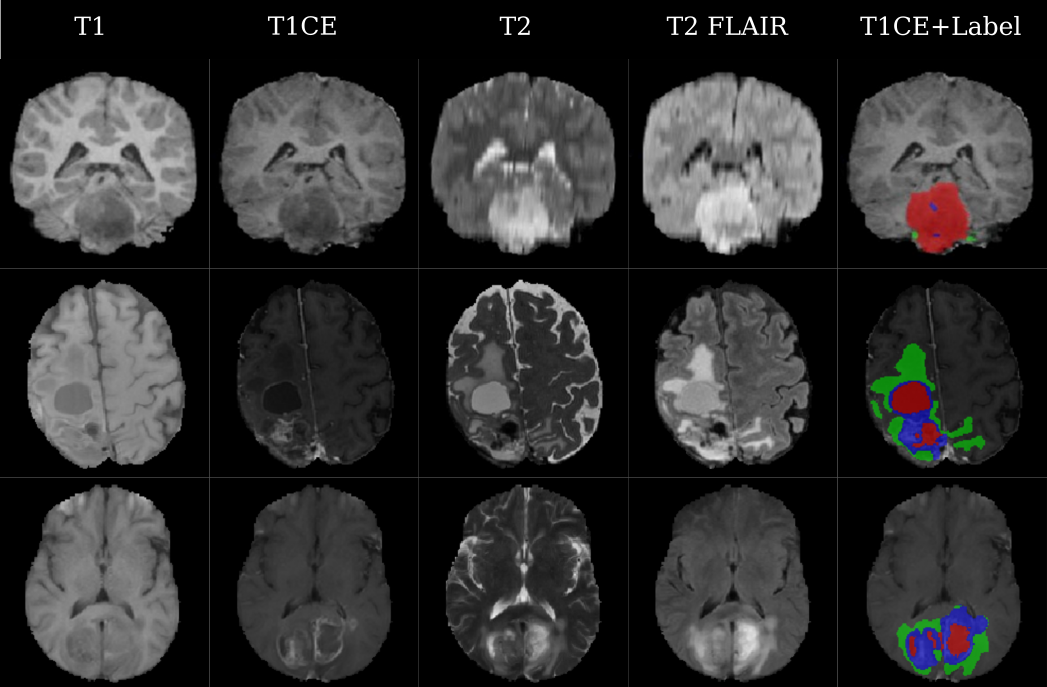}
  \caption{Training examples in the PED, MEN, and MET tasks (from top to bottom)  with the following tumor subregions: enhancing tumor ET (blue), a combination of nonenhancing tumor, cystic component, and necrosis NCR (red), and peritumoral edematous area ED in PED or surrounding nonenhancing FLAIR hyperintensity (SNFH) in MEN and MET (green).}
  \label{fig:training-examples}
\end{figure}

For all three sub-challenges, mpMRI data included pre- and post-gadolinium T1-weighted (T1 and T1CE), T2-weighted (T2), and T2-weighted fluid attenuated inversion recovery (T2-FLAIR) MRI. The mpMRI scans were pre-processed in a standardized fashion, including co-registration to the same anatomical template \cite{Rohlfing2010}, re-sampling to an isotropic resolution, and skull-stripping. Each task provided three manual segmentation labels with slightly different definitions. For example, the labels for PED included enhancing tumor (ET), non-enhancing component (NCR - a combination of non-enhancing tumor, cystic component, and necrosis), and peritumoral edematous area (ED). The labels of MEN and MET included enhancing tumor (ET), non-enhancing tumor core (NETC), and surrounding non-enhancing FLAIR hyperintensity (SNFH). Based on the three labels, three tumor sub-regions were defined to evaluate algorithm performance: enhancing tumor (ET), tumor coure (TC) and whole tumor (WT). The TC region included ET and NCR labels, and WT included all three labels. Fig. \ref{fig:training-examples} shows samples of training examples for the three tasks.

Imaging data for PED are split into training (n=99), validation (n=45), and additional testing  subsets. Imaging data for MEN are split into training (n=1,000), validation (n=141), and testing datasets\cite{MENarxiv}. Imaging data for MET are split into training (n=165), validation (n=31), and testing datasets \cite{METarxiv}.

\subsection{Deep Learning Models and Experimental Setup}

\subsubsection{nnU-Net}

The nnU-Net, which is based on the U-Net architecture \cite{unet}, is a self-configuring deep learning framework for semantic segmentation. According to the specific imaging modality and unique attributes of each dataset, the framework autonomously adjusts its internal configurations \cite{nnunet}. This results in an improved segmentation performance and generalization when compared to other state-of-the-art methods for biomedical image segmentation.

For each of the three tasks, we trained a full-resolution 3D nnU-Net (v2) model using a five-fold cross-validation approach. A preprocessing consisting of a zero mean unit variance normalization was applied to input images. Each input image was divided into patches of 128x128x128 voxels for PED and 128x160x112 for MEN and MET. The model output consisted on three channels corresponding to the three tumor sub-regions. Region-based training was employed and the patch size was determined by the GPU memory allocation, favoring larger patches while remaining within the GPU's capacity \cite{nnunet}. We used a class-weight loss function that combined Dice loss and cross entropy loss. To optimize the loss function, we used the stochastic gradient descent (SGD) optimizer with Nesterov momentum with the following parameters: initial learning rate of 0.01, momentum of 0.99, and weight decay of 3e-05. Each of the five folds was trained for 100 epochs on an NVIDIA A100 (40 GB) GPU. At inference time, images were predicted using a sliding window approach. The window size matched the patch size used during training. The nnU-Net implementation is available in an open-source repository: \href{https://github.com/MIC-DKFZ/nnUNet}{https://github.com/MIC-DKFZ/nnUNet}.

\subsubsection{Swin UNETR}

The Swin UNETR framework employs a vision transformer (ViT)-based \cite{vit} hierarchical structure for localized self-attention using non-overlapping windows \cite{swinunetr2,unetr,swinunetr}. Swin UNETR's innovative local window self-attention outperforms traditional ViT, which is well suited to multiscale tasks. The framework includes a 3D Swin transformer encoder with window-shifting for extended receptive fields, and it connects to a multiscale residual U-Net-like decoder to perform tasks like 3D medical image segmentation.

For each of the three tasks, we trained a full-resolution 3D Swin UNETR model using a five-fold cross-validation approach. A preprocessing consisting of a zero mean unit variance normalization was applied to input images. Each input image was sampled four times using patches of 96x96x96 voxels to fully utilize the GPU’s memory. The model output was 4 channels corresponding to the three labels and background. We used a class-weight loss function that combined Dice loss and focal loss. To optimize the loss function, we used the AdamW optimizer with an initial learning rate of 0.0001, momentum of 0.99, and weight decay of 3e-05. Each of the folds were trained for 600 epochs on an NVIDIA A5000 (24 GB) GPU and NVIDIA A6000 (48GB) GPU. The Swin UNETR implementation is part of the PyTorch-based framework MONAI: \footnote{\href{https://monai.io}{https://monai.io}}. Hyper-parameter optimization was carried out using Optuna \cite{optuna_2019}: \footnote{\href{https://optuna.readthedocs.io/}{optuna.readthedocs.io/}}.

\subsection{Model Ensembling}

To enhance the accuracy and robustness of the segmentation outcome, we propose a model ensembling strategy. This approach (see Fig. \ref{fig:pipeline}) involves harnessing the complementary strengths of the two models described, nnU-Net and Swin UNETR, to collectively address the task of pixel classification.

The ensembling strategy was optimized for each task based on each model's performance. For the PED task, we trained an nnU-Net solely on the ET region and ensembled its predictions with a Swin UNETR, which was trained on all labels, for ET region prediction. 
For TC and WT, we ensembled predictions from nnU-Net and Swin UNETR (as described previously), both trained on all labels. For MEN, predictions from both implementations, trained on all labels, were combined to generate ET, TC, and WT regions. Finally, for the MET task, only nnU-Net, trained on all labels, was used after experimental evaluation, as Swin UNETR showed inferior performance in this scenario.

When using a combination of nnU-Net and Swin UNETR, we ensembled the outputs (after applying the corresponding nonlinear activation functions) from each fold obtained during cross-validation ($k=5$) training of both models. This allowed us to leverage the advantages of both approaches. 

\subsection{Post-processing}

\begin{figure}[htbp]
  \includegraphics[width=\textwidth]{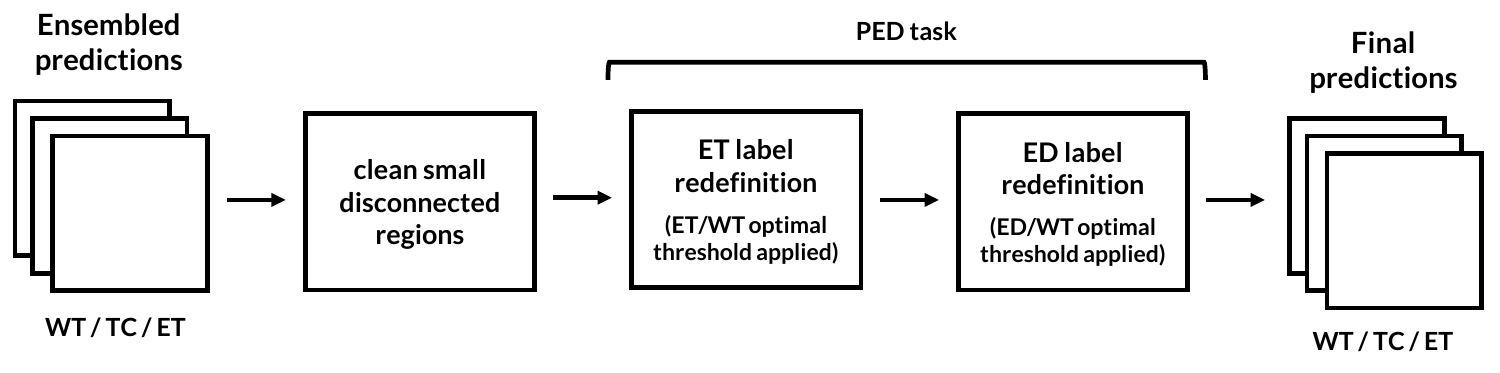}
  \caption{Post-processing strategy. The ensemble predictions were first cleaned of small disconnected regions. Then, for the PED task, ET and ED labels were redefined based on ET/WT and ED/WT thresholds, respectively.}
  \label{fig:postprocess-pipeline}
\end{figure}

\begin{figure}[htbp]
    \centering
    \includegraphics[width=\textwidth]{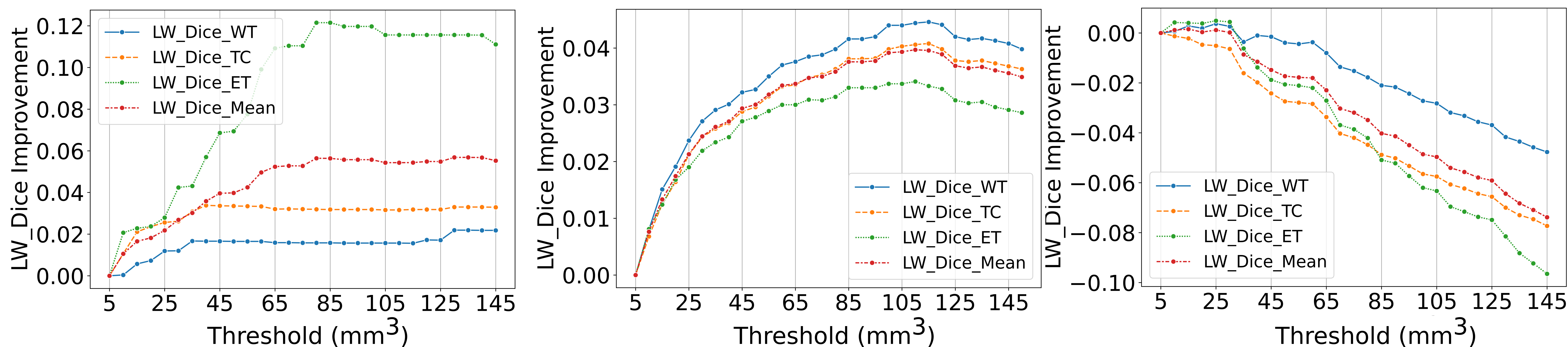}
    \caption{Threshold search on the cross-validation set for identifying small disconnected regions. LW refers to lesion-wise metrics.}
    \label{fig:cc_threshold}
\end{figure}

New performance metrics were introduced in this year’s BraTS challenge to assess segmentation models at a lesion-wise level rather than over the entire/multiple tumor region(s). We developed a post-processing strategy to adapt to the lesion-wise scores. This post-processing (see Fig.~\ref{fig:postprocess-pipeline}) was applied on the ensembled predictions and, first, removed disconnected regions (which contributed to undesired noise) smaller than 130 voxels for PED, 110 voxels for MEN, and  15 voxels for MET (Fig.~\ref{fig:cc_threshold}). These optimal threshold values were determined by experimenting with the cross-validation data.

\begin{figure}[htbp]
     \centering
     \begin{subfigure}[b]{0.49\textwidth}
         \centering
         \includegraphics[width=\textwidth]{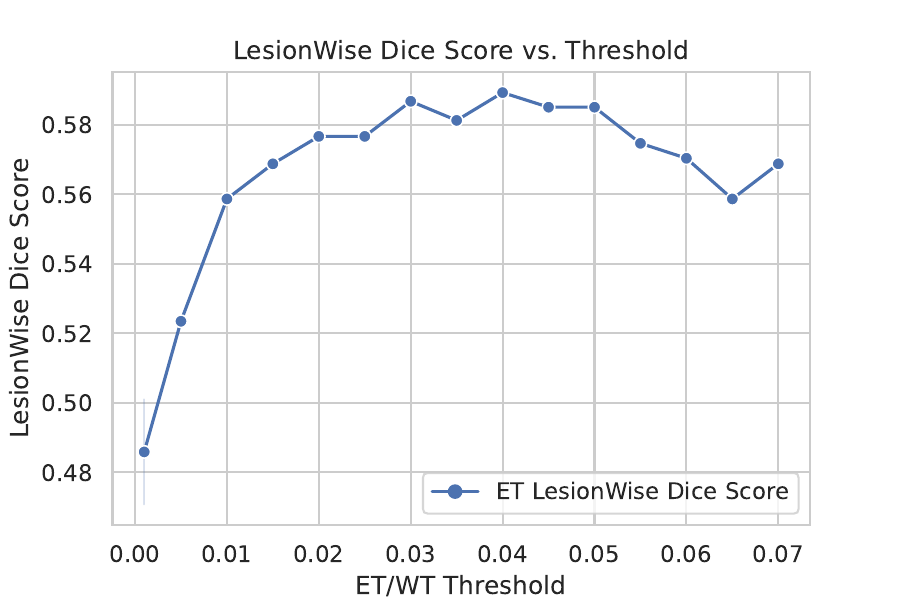}
         \caption{ET/WT threshold search}
         \label{fig:et-wt-threshold}
     \end{subfigure}
     \hfill
     \begin{subfigure}[b]{0.49\textwidth}
         \centering
         \includegraphics[width=\textwidth]{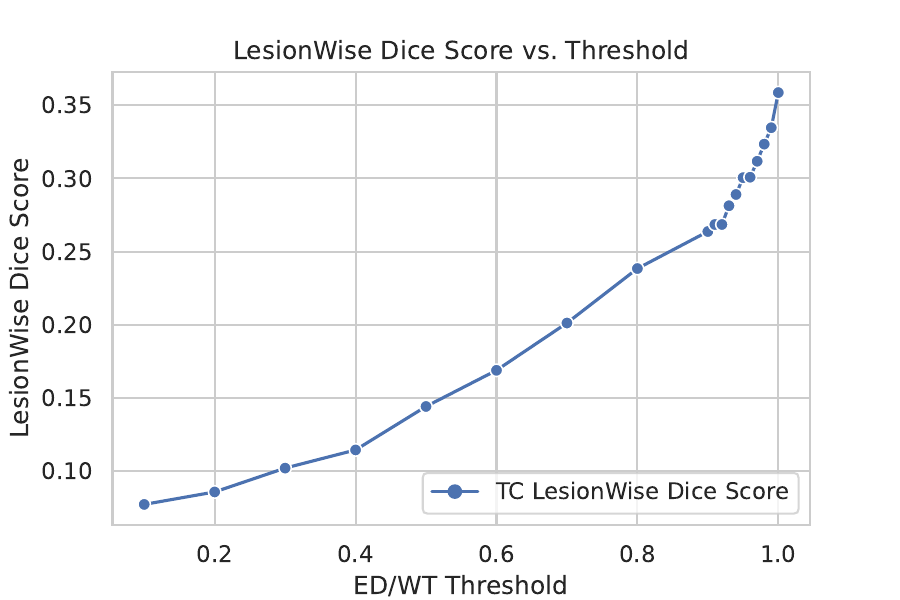}
         \caption{ED/WT threshold search}
         \label{fig:ed-wt-threshold}
     \end{subfigure}
     \caption{Threshold search on the cross-validation set. Abbreviations: ED, Peritumoral edematous; ET, enhancing tumor; WT, whole tumor.}
\end{figure}

Subsequently, given the nature of pediatric brain tumors, in the PED task, numerous cases had empty ground truth annotations for the ET and the ED labels. Therefore, redefining these labels in cases where they fell below a certain threshold with respect to the WT volume was particularly useful. For example, if the ET/WT ratio fell below the threshold, the ET label would be redefined to either NCR or ED, which corresponds to the TC region. Fig.~\ref{fig:et-wt-threshold} displays the lesion-wise Dice score \textit{vs.} ET/WT threshold curve obtained from the optimal threshold search process performed on the cross-validation sets. For ET, we obtained an optimal threshold of 0.04. On the other hand, Fig.~\ref{fig:ed-wt-threshold} displays the lesion-wise Dice score vs ED/WT threshold curve obtained from the same process but applied to the ED label, which yielded an optimal threshold of 1.00. These optimal threshold values were also determined by experimenting with the cross-validation data.

By carrying out these post-processing steps, we aimed for a balance between maintaining the integrity of the original predictions and enhancing the final scores by mitigating potential misclassifications.

\section{Results}

\begin{figure}[htbp]
    \centering
  \includegraphics[width=0.9\textwidth]{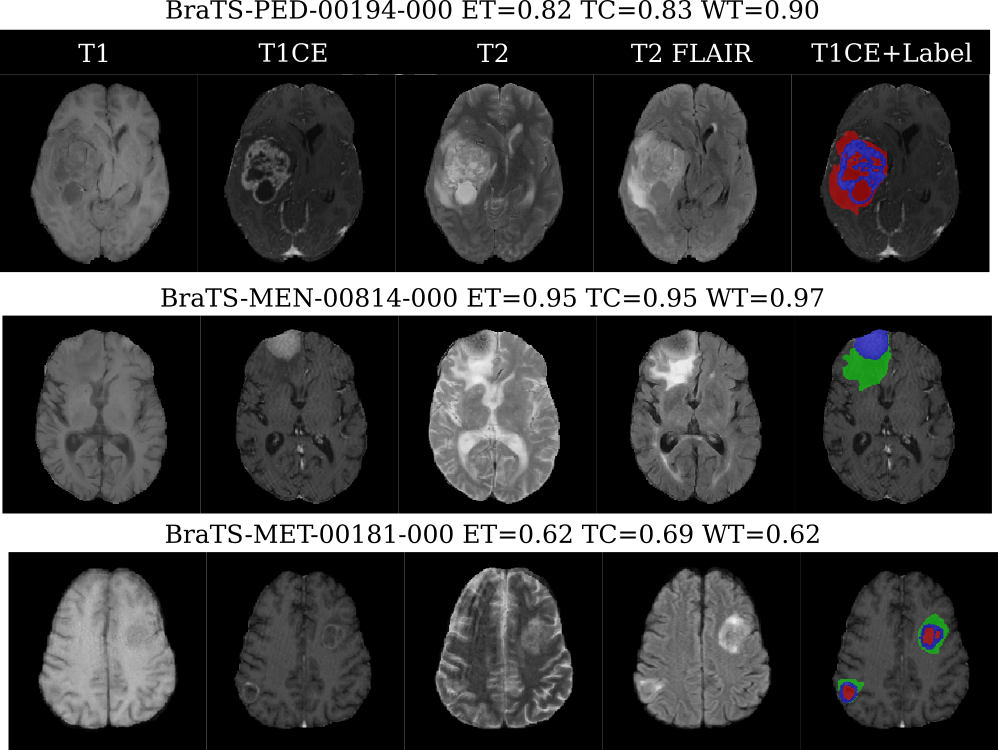}
  \caption{Qualitative results on the validation datasets of PED, MEN, and MET. The selected cases show the median of the averaged lesion-wise Dice over three tumor regions for each task (NCR-red, ED-green, ET-blue).}
  \label{fig:qual-results}
\end{figure}

\begin{table}[htbp]
\centering
\begin{tabular}{@{}clllccclccc@{}}
\toprule
\multirow{2}{*}{\textbf{Task}} &  & \multicolumn{1}{c}{\multirow{2}{*}{\textbf{Model}}} & \textbf{} & \multicolumn{3}{c}{\textbf{LW Dice}} & \textbf{} & \multicolumn{3}{c}{\textbf{LW HD95 (mm)}} \\ \cmidrule(l){4-11} 
 &  & \multicolumn{1}{c}{} & \textbf{} & \textbf{ET} & \textbf{TC} & \textbf{WT} & \textbf{} & \textbf{ET} & \textbf{TC} & \textbf{WT} \\ \midrule
\multirow{4}{*}{\textbf{PED}} &  & nnU-Net &  & 0.462 & 0.731 & 0.798 &  & 224.71 & 30.41 & 26.08 \\
 &  & Swin UNETR &  & 0.362 & 0.641 & 0.716 &  & 191.47 & 82.51 & 63.33 \\
 &  & Ensemble &  & 0.466 & 0.73 & 0.797 &  & 158.89 & 39.34 & 35.11 \\
 &  & Post-processing &  & \textbf{0.733} & \textbf{0.782} & \textbf{0.817} & \textbf{} & \textbf{75.93} & \textbf{25.54} & \textbf{24.18} \\ \midrule
\multirow{4}{*}{\textbf{MEN}} &  & nnU-Net &  & 0.818 & 0.799 & 0.794 &  & 51.06 & 55.74 & 55.59 \\
 &  & Swin UNETR &  & 0.64 & 0.644 & 0.59 &  & 117.85 & 114.43 & 135.63 \\
 &  & Ensemble &  & 0.833 & 0.832 & 0.804 &  & 45.94 & 43.68 & 53.26 \\
 &  & Post-processing &  & \textbf{0.852} & \textbf{0.846} & \textbf{0.832} & \textbf{} & \textbf{37.98} & \textbf{38.68} & \textbf{42.9} \\ \midrule
\multirow{5}{*}{\textbf{MET}} &  & nnU-Net &  & 0.565 & 0.614 & 0.545 &  & 117.53 & 110.88 & 115.5 \\
 &  & Swin UNETR &  & 0.341 & 0.378 & 0.343 &  & 204.33 & 191.12 & 200.05 \\
 &  & Ensemble &  & 0.533 & 0.594 & 0.539 &  & 123.18 & 110.28 & 115.38 \\
 &  & Post-processing &  & 0.559 & 0.604 & 0.565 &  & 110.07 & 105.12 & 102.77 \\
 &  & Without Swin &  & \textbf{0.608} & \textbf{0.649} & \textbf{0.587} &  & \textbf{91.62} & \textbf{91.29} & \textbf{95.7} \\ \bottomrule
\end{tabular}%
\vspace{0.2cm}
\caption{Quantitative results on the validation datasets of PED, MEN and MET. Lesion-wise (LW) Dice coefficients and 95\% Hausdorff distances (HD95) were computed for enhancing tumor (ET), tumor core (TC), and whole tumor (WT), respectively. Numbers in bold indicate the best results for each task.}
\label{tab:val-results}
\end{table}
Table \ref{tab:val-results} provides a comprehensive overview of the performance evaluation of our models across the validation datasets for each task. This performance evaluation was performed automatically by the challenge's digital platform, with no access to the validation ground truth data. Additionally, Fig.~\ref{fig:qual-results} shows qualitative results on some cases of the validation datasets for PED, MEN, and MET tasks.

Table \ref{tab:test-results} provides quantitative evaluation of our proposed solution across the test datasets for each task. 

\begin{table}[htb]
\centering
\begin{tabular}{@{}llllccccccc@{}}
\toprule
\multirow{2}{*}{\textbf{Task}} &  & \multicolumn{1}{c}{\multirow{2}{*}{\textbf{Statistic}}} &  & \multicolumn{3}{c}{\textbf{LW Dice}} & \multicolumn{1}{l}{} & \multicolumn{3}{c}{\textbf{LW HD95 (mm)}} \\ \cmidrule(l){5-11} 
 &  & \multicolumn{1}{c}{} &  & \textbf{ET} & \textbf{TC} & \textbf{WT} & \multicolumn{1}{l}{\textbf{}} & \textbf{ET} & \textbf{TC} & \textbf{WT} \\ \midrule
\multirow{2}{*}{\textbf{PED}} &  & \textbf{Mean} &  & 0.653 & 0.809 & 0.826 &  & 43.89 & 21.82 & 20.86 \\
&  & \textbf{(Standard deviation)} &  & (0.32) & (0.185) &  (0.183) &  & (108.59) & (75.15) & (75.31) \\
 \multirow{2}{*}{\textbf{1st rank}} &  & \textbf{25th quantile} &  & 0.569 & 0.809 & 0.840 &  & 1.41 & 2.96 & 2.96 \\
 &  & \textbf{Median} &  & 0.741 & 0.856 & 0.875 &  & 3.67 & 5.26 & 4.24 \\
 &  & \textbf{75th quantile} &  & 0.916 & 0.888 & 0.895 &  & 8.98 & 9.17 & 6.76 \\ \midrule
\multirow{2}{*}{\textbf{MEN}} &  & \textbf{Mean} &  & 0.876 & 0.867 & 0.849 &  & 30.04 & 31.69 & 35.17 \\
&  & \textbf{(Standard deviation)} &  & (0.217) & (0.227) & (0.231) &  & (80.9) & (83.53) & (86.77) \\
 \multirow{2}{*}{\textbf{3rd rank}} &  & \textbf{25th quantile} &  & 0.907 & 0.885 & 0.863 &  & 1.00 & 1.00 & 1.00 \\
 &  & \textbf{Median} &  & 0.968 & 0.968 & 0.953 &  & 1.00 & 1.00 & 1.62 \\
 &  & \textbf{75th quantile} &  & 0.985 & 0.985 & 0.975 &  & 2.91 & 3.61 & 4.36 \\ \midrule
\multirow{2}{*}{\textbf{MET}} &  & \textbf{Mean} &  & 0.555 & 0.6 & 0.58 &  & 113.96 & 112.84 & 108.17 \\
&  & \textbf{(Standard deviation)} &  & (0.279) & (0.3) & (0.289) &  & (122.53) & (121.6) & (122.13) \\
 \multirow{2}{*}{\textbf{4th rank}} &  & \textbf{25th quantile} &  & 0.312 & 0.370 & 0.342 &  & 1.66 & 1.29 & 2.45 \\
 &  & \textbf{Median} &  & 0.638 & 0.691 & 0.644 &  & 75.78 & 75.60 & 9.49 \\
 &  & \textbf{75th quantile} &  & 0.813 & 0.863 & 0.826 &  & 202.39 & 189.43 & 189.03 \\ \bottomrule
\end{tabular}%
\vspace{0.2cm}
\caption{Quantitative results on the testing datasets of PED, MEN and MET and ranks obtained in the competition. Lesion-wise (LW) Dice coefficients and 95\% Hausdorff distances (HD95) were computed for enhancing tumor (ET), tumor core (TC), and whole tumor (WT), respectively.}
\vspace{-0.5cm}
\label{tab:test-results}
\end{table}

\section{Discussion}

Overall, the combination of ensemble models and post-processing procedures yielded clear improvements in the performance for the PED, MEN, and MET tasks. With our approach, we are able to synergize the capabilities of nnU-Net and Swin UNETR models based on the validation scores to effectively generate more accurate and robust segmentation.  Our strategy minimized undesired smaller contributions, which would negatively impact lesion-wise performance.

Despite the consistent labeling of the three tumor sub-regions (ET, TC, WT) across all tasks, the lesions' morphology and spatial location are variable across subjects and tasks. Among these three sub-regions, ET identification was the most challenging in the PED task, as evidenced by the lowest Dice coefficient for this region. For the MEN and MET tasks, a better balance between label segmentation was achieved. This confirms the importance of applying label-wise post-processing techniques tailored to each task. Consequently, binary metrics disproportionately penalize false positive predictions. Moreover, the novel metrics centered on lesions result in severe penalties for isolated volumes that do not correspond to actual lesions. All of this accentuates the necessity for post-processing strategies to address disconnected and small components.

Each task's training process was restricted to the usage of task-specific data, adhering to the predefined guidelines of the BraTS 2023 challenge. Outside of the scope of this challenge, it would be beneficial to enhance the model's capacity with additional pre-training to ensure it generalizes effectively. 

Our intention was to tackle the diversity in the challenge data from both the perspective of the model and that of the data. Towards this, we argue that self-supervised learning strategies and the concept of foundation model \cite{cxr,dicom} holds the potential to augment model's generalizability, thus making it suitable for a more extensive range of clinical applications. The PED and MET tasks were conducted with limited available data, which is not surprising given the rarity of brain tumors. To overcome this limitation, the incorporation of generative artificial intelligence could also be a potential solution to augment the training set with synthetic data.

\section{Conclusion}

The segmentation of rare tumors from radiology images using deep learning techniques is challenging given the limited available data. We developed a model ensemble technique to address the segmentation of three types of brain tumors included in the BraTS 2023 challenge: i.e., pediatric, meningioma, and metastases. The ensemble prediction from nnU-net and Swin UNETR provided better segmentation accuracy for all tumor types. Furthermore, our targeted post-processing strategy based on cross-validated thresholding search improved the performance in all tasks, being more noticeable in pediatric tumor segmentation due to the label-wise targeted post-processing steps. Through this work, we explored ensembling techniques that leverage the generalizability of deep learning models and strategies across diverse data distributions and tasks. The success of our method is demonstrated by the challenge ranking positions on unseen test cases, being the winner in PED task, ranked third for MEN, and fourth for MET task.

\subsubsection{Acknowledgements} Partial support for this work was provided by the National Cancer Institute (UG3 CA236536) and by the Spanish  Ministerio de Ciencia e Innovación, the Agencia Estatal de Investigación and NextGenerationEU funds, under grants PDC2022-133865-I00 and PID2022-141493OB-I00, and EUCAIM project co-funded by the European Union (Grant Agreement \#101100633). The authors gratefully acknowledge the Universidad Politécnica de Madrid (www.upm.es) for providing computing resources on Magerit Supercomputer.

%
%
%
\bibliographystyle{splncs04}
\bibliography{references}
%




\end{document}